\documentclass[a4paper,accepted=2019-10-08]{quantumarticle}

\pdfoutput=1

\usepackage[utf8]{inputenc}
\usepackage[english]{babel}
\usepackage[T1]{fontenc}

\usepackage{graphicx}
\usepackage{amsmath}
\usepackage{amsfonts}
\usepackage{amssymb}
\usepackage{amsthm}

\usepackage{dsfont}
\usepackage{microtype}

\usepackage[numbers,sort&compress]{natbib}

\usepackage[breaklinks=true,colorlinks]{hyperref}

\newcommand{\coloneq}{\mathrel{\mathop:}=}

\newcommand{\dd}{\mathrm{d}}
\newcommand{\bd}{\begin{equation*}}
\newcommand{\ed}{\end{equation*}}

\newcommand{\Tr}{\operatorname{Tr}}

\newcommand{\imagt}{\operatorname{Im}}

\newcommand{\var}{\operatorname{Var}}

\newcommand{\ket}[1]{\left|{#1}\right\rangle}

\newcommand{\ketbra}[2]{\left|{#1}\middle\rangle\middle\langle{#2}\right|}
\newcommand{\proj}[1]{\ketbra{#1}{#1}}

\newcommand{\ew}[1]{\left\langle{#1}\right\rangle}
\newcommand{\kB}{k_\mathrm{B}}
\newcommand{\indexc}{\mathrm{c}}
\newcommand{\indexh}{\mathrm{h}}
\newcommand{\gc}{\gamma_\indexc}
\newcommand{\gh}{\gamma_\indexh}
\newcommand{\nc}{n_\indexc}
\newcommand{\nh}{n_\indexh}
\newcommand{\indexss}{\mathrm{ss}}
\newcommand{\indexf}{\mathrm{f}}
\newcommand{\Tc}{T_\indexc}
\newcommand{\Th}{T_\indexh}
\newcommand{\omegac}{\omega_\mathrm{c}}
\newcommand{\omegah}{\omega_\mathrm{h}}
\newcommand{\omegaf}{\omega_\mathrm{f}}
\newcommand{\sminus}{\sigma_-}
\newcommand{\splus}{\sigma_+}
\newcommand{\HJC}{H_\mathrm{JC}}

\newcommand{\Jh}{J_\indexh}

\begin{document}

\title{Concepts of work in autonomous quantum heat engines}

\author{Wolfgang Niedenzu}
\email{Wolfgang.Niedenzu@uibk.ac.at}
\affiliation{Institut f\"ur Theoretische Physik, Universit\"at Innsbruck, Technikerstra{\ss}e~21a, A-6020~Innsbruck, Austria}
\orcid{0000-0001-7122-3330}

\author{Marcus Huber}
\affiliation{Institut f\"ur Quantenoptik und Quanteninformation der \"Osterreichischen Akademie der Wissenschaften, Boltzmanngasse~3, A-1090 Vienna, Austria}
\orcid{0000-0003-1985-4623}

\author{Erez Boukobza}
\affiliation{School of Chemistry, Tel Aviv University, Tel~Aviv~6997801, Israel}
\affiliation{Chemistry Department, Nuclear Research Center Negev, Israel}

\date{10 October 2019}

\begin{abstract}
  One of the fundamental questions in quantum thermodynamics concerns the decomposition of energetic changes into heat and work. Contrary to classical engines, the entropy change of the piston cannot be neglected in the quantum domain. As a consequence, different concepts of work arise, depending on the desired task and the implied capabilities of the agent using the work generated by the engine. Each work quantifier---from ergotropy to non-equilibrium free energy---has well defined operational interpretations. We analyse these work quantifiers for a heat-pumped three-level maser and derive the respective engine efficiencies. In the classical limit of strong maser intensities the engine efficiency converges towards the Scovil--Schulz-DuBois maser efficiency, irrespective of the work quantifier.
\end{abstract}

\maketitle

\section{Introduction and motivation}

The question of what is \emph{quantum} in quantum heat engines (QHEs) exists since the advent of the field of quantum thermodynamics~\cite{scovil1959three,alicki1979quantum,kosloff1984quantum}. Naturally, the answer to this question requires a comparison with classical heat engines. Basically, a heat engine is a machine that converts thermal energy into work, irrespective whether its constituents are of classical or quantum nature. Classically, there exists an unambiguous notion of ``work'' and heat engines are commonly studied by analysing idealised thermodynamic cycles (comparative processes), without specifying the details of the work extraction mechanism~\cite{callenbook,cengelbook}. The underlying assumption for treating the entire energy exchanged between the engine's working medium and the piston as work is that the entropy of the work extraction device (i.e., the piston) remains constant. The produced work is further assumed to be immediately transferred to a load such that, mathematically, heat engines can be described by a periodic, time-dependent (controlled) Hamiltonian.

\par

The concept of a working medium undergoing a prescribed thermodynamic engine cycle has been very successfully applied in the quantum domain too, both theoretically~\cite{alicki1979quantum,kosloff1984quantum,geva1992quantum,kosloff2013quantum,gelbwaser2015thermodynamics,vinjanampathy2016quantum,kosloff2017quantum,binder2019thermodynamicsbook} and experimentally~\cite{koski2014experimental,rossnagel2016single,klaers2017squeezed,vanhorne2018single,klatzow2019experimental}. Hereby the macroscopic working medium (e.g., an air-fuel mixture) is replaced by a quantum system, e.g., a single spin or a single atom. The work extraction mechanism, by contrast, is considered to be classical with a driving field being the analogue of a mechanical piston. Therefore, the unambiguous notion of work from classical thermodynamics, namely, the energetic change of this field (piston), also applies here~\cite{alicki1979quantum,kosloff1984quantum}.

\par
\begin{figure}
  \centering
  \includegraphics[width=\columnwidth]{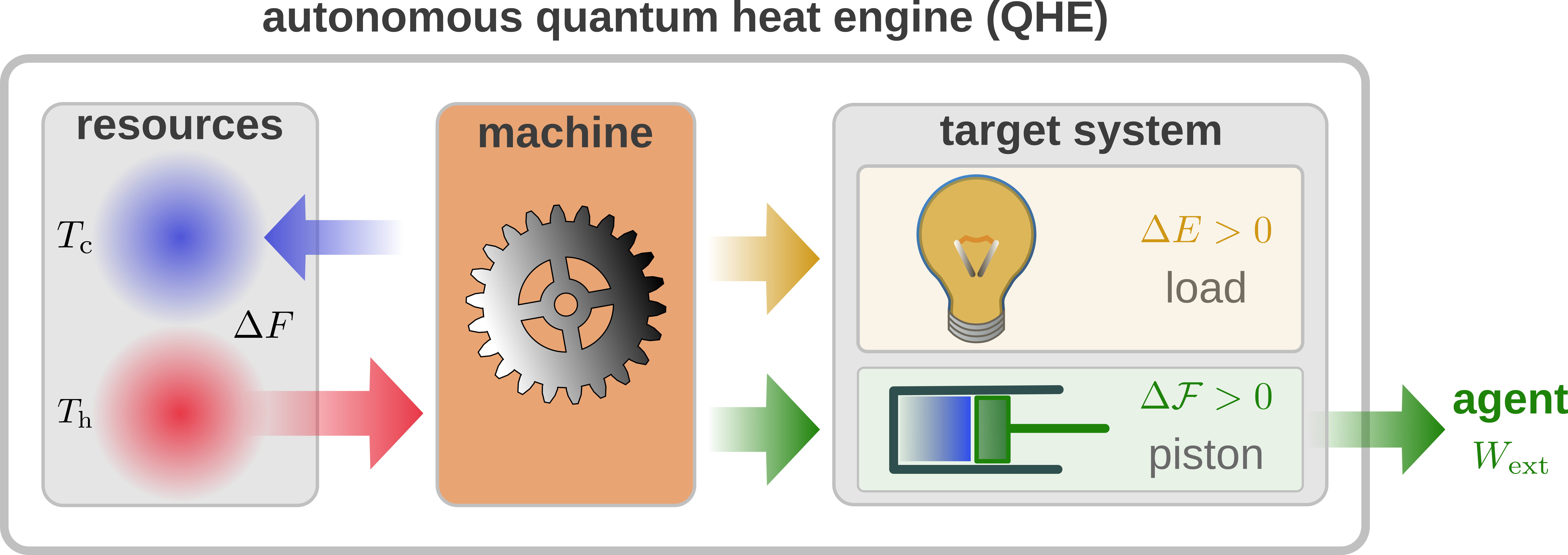}
  \caption{An autonomous quantum heat engine (QHE) uses the equilibrium free energy difference $\Delta F$ between two thermal baths at temperatures $\Tc$ and $\Th$, respectively, to autonomously drive a second quantum system into a specific quantum state. The target system thereby either constitutes a load of the engine (yellow branch), such that its energy increase $\Delta E$ matters, or plays the role of a piston (green branch), such that its non-equilibrium free energy increase $\Delta \mathcal{F}$ matters. In the latter case the piston's non-equilibrium free energy is subsequently used by an agent to perform external work $W_\mathrm{ext}$ in a controlled (non-autonomous) process.}\label{fig_engine}
\end{figure}
\par

An externally prescribed periodic engine cycle is of course an idealisation. Instead of being externally-controlled, one may include the piston degrees of freedom into the dynamics by considering a time-independent (autonomous) Hamiltonian for the joint working-medium--piston system. The different strokes of the underlying thermodynamic cycle are then triggered by, e.g., the piston position~\cite{tonner2005autonomous,roulet2017autonomous} rather than by an external control field. Such self-contained heat engines typically autonomously amplify the energy of a prescribed initial state of the piston subsystem~\cite{gelbwaser2013work,gelbwaser2014heat,levy2016quantum,roulet2017autonomous,ghosh2018two}; this initial state needs to be provided by an external agent. Contrary to driven heat engines the work is accumulated in the piston and causes the continuous amplification (e.g., acceleration) of the piston motion if the work performed by the engine is not further unidirectionaly transferred to a load~\cite{levy2016quantum,teo2017converting,seah2018work}. For this reason and also due to the piston entropy not remaining constant any more, the engine no longer operates in a cyclic fashion. Also, the initial amplification of an input state is often analysed in the limit of short interaction times, permitting a separation ansatz~\cite{gelbwaser2013work,gelbwaser2014heat,ghosh2018two}.

\par

Hence, while these engines autonomously convert heat to work, without the need of an externally-prescribed cycle, they nevertheless require an external agent to initialise their respective input states. Being an external out-of-equilibrium resource~\cite{mari2015quantum}, the latter has a fundamental influence on the engine operation---if it is unfavourably chosen, the thermodynamic machine may not act as an engine and only heat up the piston~\cite{gelbwaser2013work,gelbwaser2014heat}. Such self-contained engines may be dubbed \emph{quantised} heat engines to stress that whilst their constituents may be quantum, their operational principle is still conceptually of classical origin, e.g., based on the concept that externally changing a parameter requires a different amount of energy depending on the temperature of the body. In that sense it is not of conceptual importance whether this parameter is the volume of a gas or the energetic gap between the discrete energy levels of a two-level atom. Such a self-contained heat engine has recently been experimentally realised based on a single spin~\cite{vonlindenfels2019spin}.

\par

The analysis of autonomous quantised heat engines sparked an ongoing debate on the nature of work in autonomous quantum setups~\cite{pusz1978passive,lenard1978thermodynamical,allahverdyan2004maximal,deffner2008nonequilibrium,dahlsten2011inadequacy,alicki2013entanglement,dorner2013extracting,gelbwaser2013work,hovhannisyan2013entanglement,gelbwaser2014heat,skrzypczyk2014work,elouard2015reversible,perarnau2015extractable,brown2016passivity,gallego2016thermodynamic,horowitz2016work,korzekwa2016extraction,talkner2016aspects,ghosh2018two,loerch2018optimal,seah2018work,baeumer2019fluctuating,tobalina2019vanishing}. Although this debate is not yet settled, the concept of ergotropy~\cite{pusz1978passive,lenard1978thermodynamical,allahverdyan2004maximal} being a quantum analogue of work for the considered tasks has gathered strong support in the quantum thermodynamics community. Loosely speaking, ergotropy is that part of the energy of a quantum system that can be extracted in a unitary (and therefore isentropic) fashion by an agent. According to this view, an engine increases the ergotropy of the piston mode. The remainder of the transferred energy is then of thermal nature and heats up the piston mode~\cite{gelbwaser2013work,gelbwaser2014heat,ghosh2018two}. While the entropy associated to this thermal energy may be considerable for small quantum systems, in the classical limit it hardly contributes to the total piston energy such that the entire energy transfer may be viewed as contributing to ergotropy, which is also defined for classical systems~\cite{gorecki1980passive,daniels1981passivity,daprovidencia1987variational}. This justifies the analysis of this type of heat engines by means of idealised, prescribed thermodynamic cycles once the piston becomes so strongly populated that its passive (i.e., non-ergotropic) energy may be neglected. By doing this any entanglement or correlations between the engine working medium and the piston are also neglected.

\par

In principle, autonomous engines that mimic thermodynamic cycles may either be classical or quantum, depending on the engine design and its size. One may, however, also consider a conceptually and physically very distinct type of autonomous QHEs, namely those \emph{without} driven counterpart. These engines are not obtained by quantising a classical piston mode (driving field) and are thus not described by self-contained versions of time-dependent Hamiltonians. They may not amplify an externally-prescribed input state but operate under steady-state conditions, independent of the initial condition~\cite{mari2015quantum}. Hence, their operation does not require any external agent. The operational principle of such heat engines may heavily rely on the presence of quantised energy levels, quantum correlations or entanglement between its constituents and may therefore not even possess a classical counterpart. A prime example for such quantum engines are heat-pumped masers or lasers~\cite{scovil1959three,geva1996quantum,boukobza2006thermodynamic,sandner2012temperature,boukobza2013breaking,perl2017thermodynamic}. We may dub such engines \emph{quantum} heat engines to distinguish them from the quantised heat engines introduced above.

\par

Note, however, that there is no universally-accepted criterion for ``quantumness''. Indeed, whether a particular system is deemed to be ``quantum'' or not is very differently assessed depending on the field of research and application in mind and may, e.g., relate to negative quasi-probability distributions~\cite{wallsbook,niedenzu2016operation}, the presence of coherence or entanglement~\cite{friis2018entanglement} or whether a system cannot be efficiently simulated on a classical computer~\cite{preskill2018quantum}. Therefore, the purpose of the above division of autonomous engines depending on their ``quantumness'' is mainly for semantic convenience; in either case we consider few-body systems that constitute autonomous thermodynamic engines.

\par

As mentioned above, classically, the energy associated to the entropy change of the piston may be neglected and the entire transferred energy be regarded to constitute ``useful work''. Quantum-mechanically, however, the situation is more intricate. Owing to the smallness of the systems involved, the entropy of the piston is no longer negligible and may significantly hamper work extraction. At this point, however, one needs to specify the tasks of the autonomous QHE: Does the second quantum system (which is coupled to the working medium) constitute a load, from which no work is subsequently extracted, or a piston, whose quantum state is later exploited by an external agent to extract work (Fig.~\ref{fig_engine})? In the latter case one must also take this agent's abilities into account~\cite{faist2018fundamental}. Consequently, quantum-mechanically it is inevitable to specify the task of the engine in order to be able to quantify work~\cite{gallego2016thermodynamic} and efficiency. We note, however, that depending on the task of an autonomous QHE no notion of work may be necessary to assess its performance, e.g., for entanglement generation~\cite{tavakoli2018heralded}, time keeping~\cite{erker2017autonomous} or refrigeration~\cite{mitchison2015coherence}.

\par

Here we clarify the operational meaning of different measures of work for autonomous quantum heat engines and reveal the intimate relation between ergotropy and non-equilibrium free energy. We illustrate the operational meaning of these work quantifiers and exemplify them by the heat-pumped maser~\cite{geva1996quantum,boukobza2006thermodynamic,boukobza2013breaking,perl2017thermodynamic}. Work in quantum mechanics is not universal and a unified notion of work only emerges in the classical limit of strong maser intensities, where the concrete measure does not matter any more. We show that contrary to the aforementioned case of quantised heat engines with driven counterpart, in the considered QHEs the physical origin of the piston entropy may be entirely non-thermal: Rather than stemming from classical heating of the light field, it is the undetermined phase of the laser light that generates the entropy. Notwithstanding, even if in the classical limit of a large piston population (laser intensity) the work and efficiency measures reveal a unified ``classical'' behaviour, the QHE itself remains inherently quantum in its operation: The maser does not have a classical analogue and as such still relies on quantum features, e.g., entanglement and discrete energy levels, to convert heat into work.

\par

These conceptual differences make autonomous quantum heat engines (QHEs) devoid of any driven counterpart ideally suited for investigating genuine quantum phenomena in the operation of heat engines.

\section{Energetics of the piston mode}

The work produced by an autonomous QHE may be directly ``cashed in'' by a load to increase its energy while being driven into some desired quantum state (yellow branch in Fig.~\ref{fig_engine}). If, however, an agent envisages to use this state to subsequently perform work on some external system, the load now adopts the role of a piston (green branch in Fig.~\ref{fig_engine}). Then the change in the load state is not the end of the story and ambiguous definitions of external work arise. By contrast, the ``internal'' work performed by the machine on the load/piston is not measurable and hence irrelevant. Work, as introduced in classical thermodynamics, is inherently an agent-based task-related concept~\cite{jaynes1992gibbs,alicki2017gkls}. Indeed, thermodynamic concepts in the quantum regime are also often subjective and the emergence of more objective notions requires asymptotic sizes or additional assumptions. Thus, any operational approach to work will feature an agent~\cite{gallego2016thermodynamic,boes2018statistical,faist2018fundamental,bera2019thermodynamics,boes2019vonneumann,wilming2019equilibration}.

\par
\begin{figure}
  \centering
  \includegraphics[width=\columnwidth]{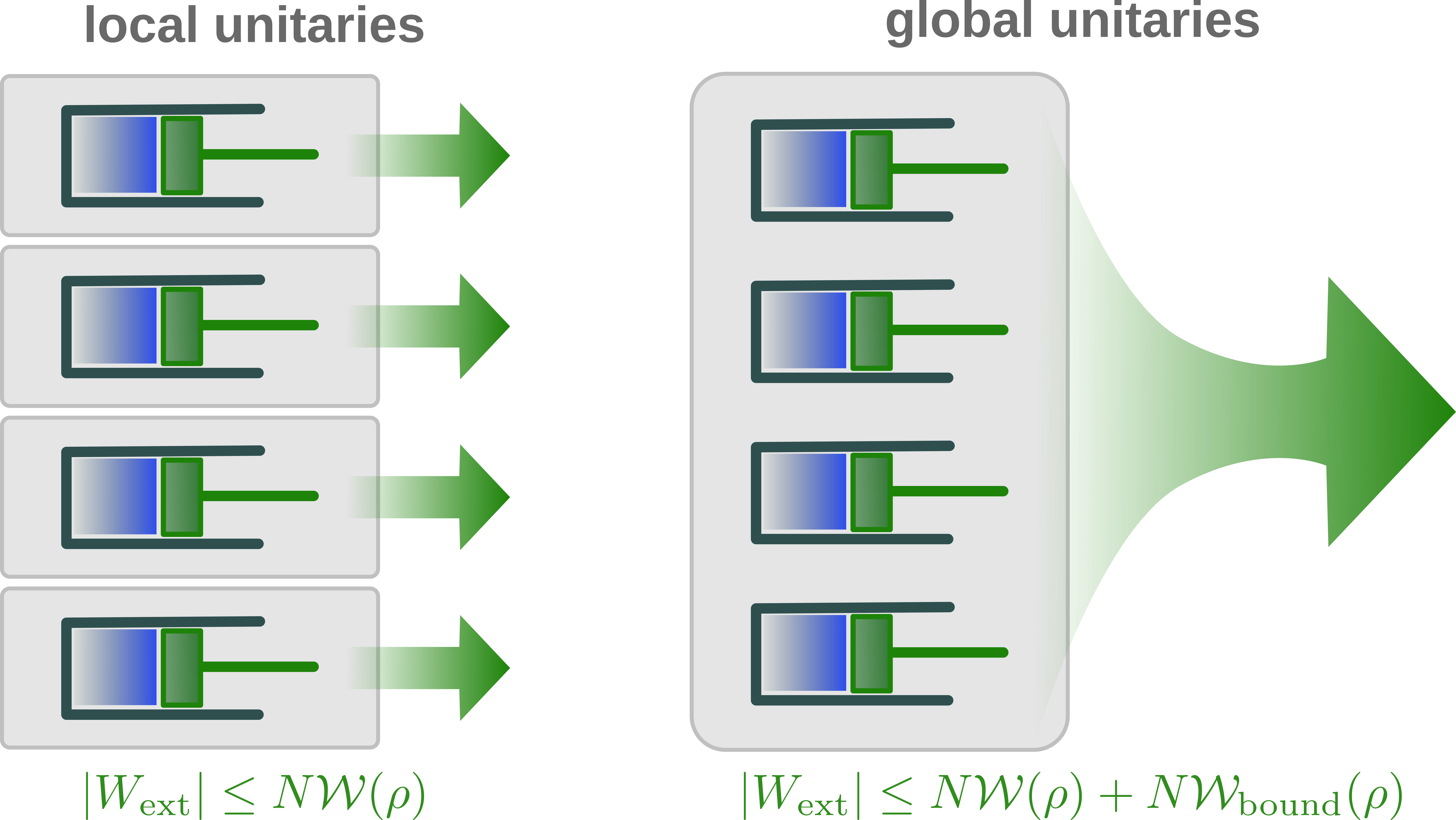}
  \caption{Extractable work from $N$ copies of the piston state $\rho$ by means of unitary transformations applied by an agent. Using local unitaries (each acting on a single copy), the agent can maximally extract the work $N\mathcal{W}(\rho)$, where $\mathcal{W}(\rho)$ is the ergotropy of a single copy of $\rho$. The bound ergotropy of each piston state is unitarily inaccessible and thus contributes to the passive energy. By contrast, if the agent is capable of applying global unitaries that act on all the copies, the bound ergotropy of every copy becomes unitarily accessible and thus enlarges the external work $|W_\mathrm{ext}|$. At most, $N\mathcal{W}(\rho)+N\mathcal{W}_\mathrm{bound}(\rho)$ can be extracted (in the limit $N\rightarrow\infty$); for finite $N$ a part of the bound ergotropy remains passive energy. Ergotropy is a non-extensive quantity and therefore the energetic weight of the piston entropy (i.e., the passive energy) is reduced as an agent increases its control capabilities on the piston ensemble.}\label{fig_piston_unitaries}
\end{figure}
\par

\begin{figure}
  \centering
  \includegraphics[width=.64\columnwidth]{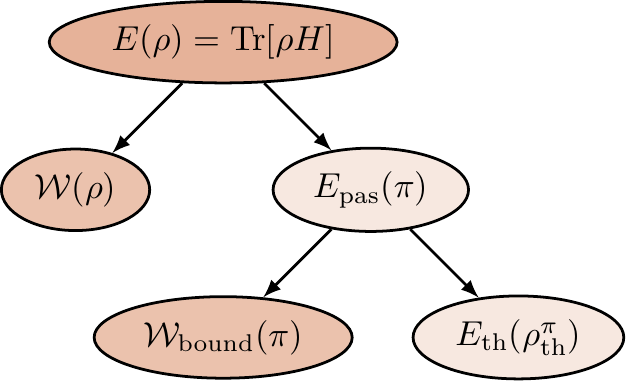}
  \caption{Visualisation of the decomposition of the energy $E$ of a quantum state $\rho$ from Eqs.~\eqref{eq_E_W_Epas}--\eqref{eq_Wtot}.}\label{fig_tree_energy}
\end{figure}

In order to understand how an external agent can make use of the piston state $\rho$, we have to further understand the energetic content of the latter. To this end we first decompose the energy $E=\Tr[\rho H]$ as
\begin{equation}\label{eq_E_W_Epas}
  E=\mathcal{W}+E_\mathrm{pas},
\end{equation}
where $\mathcal{W}$ is the ergotropy of $\rho$, i.e., the maximum energy extractable by cyclic unitaries~\cite{pusz1978passive,lenard1978thermodynamical,allahverdyan2004maximal} (a unitary applied on a system is called cyclic when the initial and final system Hamiltonians coincide). The remaining energy $E_\mathrm{pas}=\Tr[\pi H]$ that is not accessible by such unitaries is attributed to the passive state $\pi$ of $\rho$, to which it is unitarily related, $\rho=U\pi U^\dagger$. This passive state, however, is not necessarily \emph{completely passive}. Namely, if $\pi$ is not a Gibbs state then the energy of a collection of $N$ copies of $\pi$ can be further reduced by global cyclic unitaries that act on all the $N$ copies~\cite{lenard1978thermodynamical}. In other words, contrary to energy, ergotropy is, in general, a \emph{non-extensive} quantity, $\mathcal{W}(\rho^{\otimes N})\geq N\mathcal{W}(\rho)$, except if $\pi$ is a thermal state (equal sign). We therefore further decompose the energy~\eqref{eq_E_W_Epas} as
\begin{equation}\label{eq_E_W_Wbound_Eth}
  E=\mathcal{W}+\mathcal{W}_\mathrm{bound}+E_\mathrm{th},
\end{equation}
where
\begin{equation}\label{eq_W_bound}
  \mathcal{W}_\mathrm{bound}\coloneq E_\mathrm{pas}-E_\mathrm{th}
\end{equation}
is ``quantum-bound ergotropy'' and $E_\mathrm{th}$ the energy of the thermal state $\rho_\mathrm{th}^\pi$ with the same entropy as $\pi$, $\mathcal{S}(\rho_\mathrm{th}^\pi)=\mathcal{S}(\pi)\equiv\mathcal{S}(\rho)$. Since thermal states are the minimum-energy states for a given entropy~\cite{schwablbook} it is guaranteed that $\mathcal{W}_\mathrm{bound}\geq0$. Hence,
\begin{equation}\label{eq_Wtot}
  \mathcal{W}_\mathrm{tot}\coloneq\mathcal{W}+\mathcal{W}_\mathrm{bound}
\end{equation}
is the total ergotropy that can be extracted from each copy of $\rho$ by cyclic unitaries that act on an ensemble of $N\rightarrow\infty$ copies of $\rho$ (Fig.~\ref{fig_piston_unitaries}). Only in the latter limit is the passive state of $\rho^{\otimes N}$ a (completely passive) Gibbs state and its energy therefore of purely thermal nature. We may thus think of the non-thermal part of the passive energy as bound ergotropy. The energetic hierarchy given by Eqs.~\eqref{eq_E_W_Epas}--\eqref{eq_Wtot} is shown in Fig.~\ref{fig_tree_energy}. Note that ergotropy extraction, i.e., unitary energy reduction, is the widely-accepted notion of work in driven quantum heat engines~\cite{alicki1979quantum,kosloff1984quantum}.

\par
\par

Equations~\eqref{eq_E_W_Wbound_Eth} and~\eqref{eq_Wtot} carry the following meaning: During the engine operation the piston energy changes by
\begin{equation}
  \Delta E = \Delta\mathcal{W}_\mathrm{tot}+\Delta E_\mathrm{th}.
\end{equation}
The actual amount of work that an external agent can readily unitarily extract from the imparted ergotropy, however, strongly depends on this agent's capabilities. For example, if the agent only has access to a single copy of $\rho$, then the bound ergotropy~\eqref{eq_W_bound} is perceived as passive energy. This energy is then an analogue of ``heat'' since it is (i)~lost for direct (unitary) work extraction and (ii)~contributes to the piston entropy. By contrast, if the agent has access to many copies of $\rho$, the bound ergotropy is ``zero-entropy energy'' and becomes unitarily accessible. The entropy $N\mathcal{S}(\rho)$ of $N$ copies of the piston state then corresponds to thermal energy and thus has a lower energetic weight than the entropy $\mathcal{S}$ of a single copy (or $N$ copies with only local unitaries), where it corresponds to the sum of thermal energy and bound ergotropy. In other words, full knowledge of the state $\rho$ and full control on the piston is required to allow an agent to extract the entire ergotropy; knowing $\rho$ but lacking control or having full control but lacking knowledge on $\rho$ always increases the passive energy perceived by the agent. In principle, the required level of control for extracting the entire bound ergotropy may be arbitrary complex and, e.g., involve several entangling operations. As a consequence, it may be formidably hard to unitarily access this energy in practical scenarios, except for special cases.

\par

\begin{figure}
  \centering
  \includegraphics[width=.68\columnwidth]{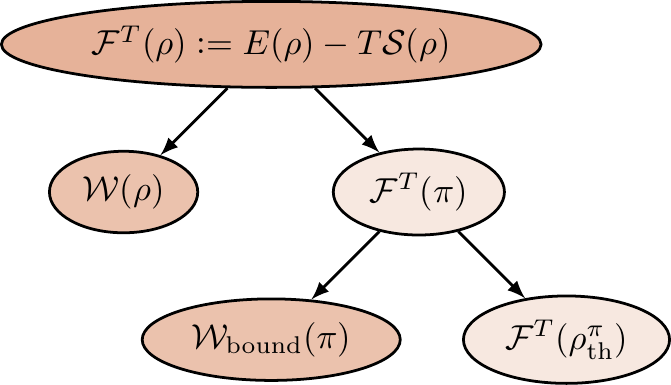}
  \caption{Visualisation of the decomposition of the non-equilibrium free energy $\mathcal{F}$ of a quantum state $\rho$ from Eqs.~\eqref{eq_free_energy_noneq}--\eqref{eq_F_erg_bound_th}.}\label{fig_tree_free_energy}
\end{figure}

Ergotropy is the readily available work, similar to a work reservoir (battery). Instead of applying cyclic unitaries the agent may, however, also generate work in a subsequent external thermodynamic process that involves a heat bath at temperature $T$ (which may be one of the two available temperatures $\Tc$ and $\Th$). Namely, the agent may use the piston's \emph{non-equilibrium} free energy w.r.t.\ $T$, defined as~\cite{esposito2011second,gardas2015thermodynamic,parrondo2015thermodynamics}
\begin{equation}\label{eq_free_energy_noneq}
  \mathcal{F}^T(\rho)\coloneq E(\rho)-T\mathcal{S}(\rho).
\end{equation}
The non-equilibrium free energy naturally relates to the concepts of ergotropy and passive energy,
\begin{equation}\label{eq_F_erg_Fpi}
  \mathcal{F}^T(\rho)= \mathcal{W}(\rho)+E_\mathrm{pas}(\pi)-T\mathcal{S}(\rho)=\mathcal{W}(\rho)+\mathcal{F}^T(\pi),
\end{equation}
where the second equal sign follows from $\mathcal{S}(\rho)\equiv\mathcal{S}(\pi)$. Namely, the non-equilibrium free energy of the state $\rho$ equals this state's ergotropy plus the non-equilibrium free energy of the passive state $\pi$. Using the notion of bound ergotropy introduced in Eq.~\eqref{eq_W_bound}, Eq.~\eqref{eq_F_erg_Fpi} may further be decomposed as
\begin{equation}\label{eq_F_erg_bound_th}
  \mathcal{F}^T(\rho)= \mathcal{W}(\rho)+\mathcal{W}_\mathrm{bound}(\pi)+\mathcal{F}^T(\rho_\mathrm{th}^\pi).
\end{equation}
The free-energy hierarchy given by Eqs.~\eqref{eq_free_energy_noneq}--\eqref{eq_F_erg_bound_th} is shown in Fig.~\ref{fig_tree_free_energy}.

\par
\par

Equations~\eqref{eq_F_erg_Fpi} and~\eqref{eq_F_erg_bound_th} close the bridge between the concepts of ergotropy and free energy: The ergotropy $\mathcal{W}(\rho)$ is the ``battery-like'' part of the free energy that can readily be extracted in the form of work. By contrast, the ergotropy of the passive state $\pi$ is bound and requires global cyclic unitaries acting on multiple copies of $\pi$ to be ``unlocked''. The remaining entropic part of the free energy is then of thermal nature. If, however, only a single copy of $\rho$ is available, then its free energy consists of ergotropy and the free energy of the passive state. The latter may then be transformed into work in a non-cyclic thermodynamic process involving a bath at temperature $T$ and as such does not constitute a work reservoir (battery).

\par

Equation~\eqref{eq_F_erg_bound_th} hence suggests the following operational interpretation for most favourably using the non-equilibrium free energy: First, a maximum of energy should be extracted in a unitary way; the concrete amount will depend depend on the available control and the number of copies of $\rho$. The free energy of the remaining passive state, which in the ideal case is a thermal state, can then be further used in a non-unitary thermodynamic process.

\section{Illustrative example: The three-level maser and its operation modes}

\begin{figure}
  \centering
  \includegraphics[width=\columnwidth]{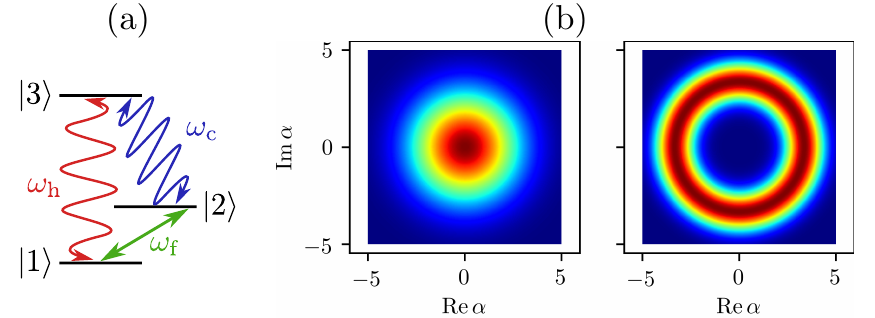}
  \caption{Heat-pumped three-level maser. (a)~The hot thermal bath at temperature $\Th$ couples the states $\ket{1}$ and $\ket{3}$ at frequency $\omegah\coloneq \omega_3-\omega_1$ whereas the cold bath at temperature $\Tc$ couples $\ket{3}$ and $\ket{2}$ at frequency $\omegac\coloneq \omega_3-\omega_2$. The lasing transition $\ket{1}\leftrightarrow\ket{2}$ of frequency $\omegaf\coloneq \omega_2-\omega_1$ is resonantly coupled to a single cavity-field mode (harmonic oscillator). (b)~$Q$-function of the cavity mode below (left) and above (right) the maser threshold, respectively. See Appendix~\ref{app_ehrenfest} for details on the numerical simulations.}\label{fig_maser}
\end{figure}

To illustrate the foregoing general considerations we now consider the well-known Scovil--Schulz-DuBois (SSD) heat-pumped three-level maser~\cite{scovil1959three} (Fig.~\ref{fig_maser}). We stress, however, that the conclusions also apply to other autonomous setups such as (spatial) temperature-gradient lasers~\cite{sandner2012temperature}, optomechanically-coupled oscillators~\cite{mari2015quantum} or machines wherein two qubits are coupled to the respective baths and a third one mediates the interaction with a harmonic piston mode. Steady-state work production without saturation effects requires an infinitely-dimensional piston or load, e.g., a harmonic oscillator, a free particle in the gravitational field, or a rotating flywheel. For finite-dimensional target systems the engine operation is always restricted to be of transient nature.

\par

The dynamics of the joint atom--cavity system is modelled by a ``local'' Lindblad master equation whose consistency with the second law of thermodynamics has been verified in numerical simulations (see Appendix~\ref{app_ehrenfest}). Depending on the involved frequencies and temperatures, this maser exhibits different operation modes (Fig.~\ref{fig_maser}b)~\cite{boukobza2008thermodynamic}:

\par
\par

(i)~For $\omegac/\Tc<\omegah/\Th$ the atom--cavity system reaches an equilibrium state devoid of any remaining energy currents. Consequently, any refrigerator or engine operation can only be of transient nature. During the evolution the reduced state of the cavity field becomes thermal with effective temperature $T_\mathrm{eff}=\omegaf/(\omegah/\Th-\omegac/\Tc)$ and energy $E_\mathrm{eff}$ (central blob in the left panel of Fig.~\ref{fig_maser}b). The concrete dynamics of course depends on the involved time scales but as a general rule of thumb an initial state with energy larger than $E_\mathrm{eff}$ may power transient refrigeration of the cold bath from its free energy. By contrast, for a lower initial energy the field's free energy w.r.t.\ the highest available bath temperature, i.e., $\Th$, starts increasing at some point during the transient dynamics. This indicates that work was performed, even though the cavity state is thermal and does not contain any (available or bound) ergotropy.

\par
\begin{figure}
  \centering
  \includegraphics[width=0.9\columnwidth]{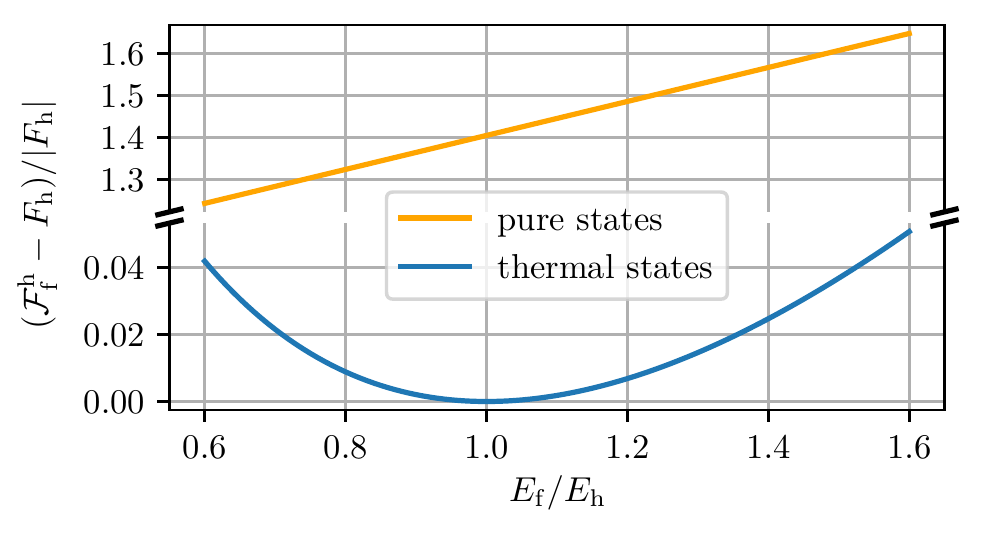}
  \caption{Non-equilibrium free energy $\mathcal{F}_\indexf^\indexh\coloneq \mathcal{F}_\indexf^{\Th}$ [Eq.~\eqref{eq_free_energy_noneq}] of the cavity field (piston) w.r.t.\ $\Th$. Thermal states with $\Tc\leq T\leq \Th$ are ``free'' resources, i.e., directly obtainable from the available heat baths without the need to build an engine. The thermal state of the cavity field at $\Th$ is then the most energetic (energy $E_\indexh$) free state and $F_\indexh$ its \emph{equilibrium} free energy. Under steady-state operation the engine performs work on the field if the field's energy and free energy relative to this state increase~\cite{mari2015quantum}. The states above the lower blue (thermal) line have a reduced entropy and thus contain ergotropy; states on the upper orange line have zero entropy and thus exclusively contain ergotropy. Since thermal states have maximum entropy for a given energy, any state of the cavity field for a given energy lies between the two (blue and orange) curves. Parameters: $\kB\Th=10\hbar\omega_\indexf$.}\label{fig_free_energy}
\end{figure}
\par

(ii)~At the masing threshold $\omegac/\Tc=\omegah/\Th$ the thermal occupations $n_i=\{\exp[\hbar\omega_i/(\kB T_i)]-1\}^{-1}$ ($i\in\{\indexc,\indexh\}$) of the two thermal baths at the two atomic transitions frequencies $\omegac$ and $\omegah$ coincide. The atom then reaches steady state while the cavity field continuously heats up, $T_\mathrm{eff}\rightarrow\infty$. Consequently, no ergotropy is accumulated in the field. Hence, contrary to below threshold the field does not reach any steady state and the machine acts as an ``eternal'' transient engine with $\dot{\mathcal{F}}^\indexh_\indexf>0$ and $\dot{\mathcal{W}}_\indexf=0$, $\dot{\mathcal{F}}^\indexh_\indexf\coloneq \dot{\mathcal{F}}_\indexf^{\Th}$ being the change in the field's non-equilibrium free energy~\eqref{eq_free_energy_noneq} w.r.t.\ $\Th$. Since the cavity field remains thermal, its non-equilibrium free energy follows the lower blue curve in Fig.~\ref{fig_free_energy}.

\par

(iii)~Above threshold, $\omegac/\Tc>\omegah/\Th$, the engine continuously performs work, $\dot{\mathcal{F}}^\indexh_\indexf>0$ and $\dot{\mathcal{W}}_\indexf>0$, but not all of the energy accumulated in the field contributes to ergotropy since also the field entropy increases. This regime of the atom attaining steady state and the field intensity growing corresponds to a steady-state operation of the engine (as opposed to the transient engine below threshold). During the engine operation the field becomes a Poissonian state (phase-averaged coherent state) of continuously increasing intensity, $\dot{E}_\indexf>0$, whose photon bunching parameter $g^{(2)}(0)\coloneq\ew{a^\dagger a^\dagger a a}/\ew{a^\dagger a}^2$ converges towards $1$ (confirmed in numerical simulations, see Appendix~\ref{app_ehrenfest}), thus revealing the Poissonian statistics~\cite{wallsbook}. Its $Q$-function has the shape of an annulus, as expected for a maser/laser~\cite{scullybook,wallsbook} (Fig.~\ref{fig_maser}b). Hence, the field's passive state is not a thermal state and contains bound ergotropy. Note that contrary to the light amplifiers considered in Refs.~\cite{gelbwaser2014heat,ghosh2018two}, the steady-state maser operates as a light generator. Here the increasing light field entropy does not stem from heating (which would cause photon bunching) but solely from the undetermined phase of the laser light. The operation above threshold may be understood as ``cashing in'' the work potential (population inversion) of the atom. Indeed, the atomic population inversion in steady state is much smaller than without coupling the atom to the cavity~\cite{boukobza2006thermodynamic}. Finally, we note that Poissonian states of the cavity field correspond to points in between the two extreme curves (thermal and pure states) in Fig.~\ref{fig_free_energy}.

\par

We note that autonomous QHEs do not operate in cycles, hence there would be no a-priori need for a cold bath. However, two thermal baths are nevertheless required to generate a continuous steady-state operation of the engine as the operation of a single-bath engine would always be of transient nature.

\par

\section{Thermodynamic tasks and efficiency of autonomous QHEs}

In the following we consider the maser as an autonomous QHE under steady-state operation (i.e., above threshold). We recall that in this regime the atomic populations do not change whereas the mean number of photons in the Poissonian light field continuously increases. One may then consider the following scenarios regarding the role of the light field:
\begin{itemize}
\item \emph{Light field as load of the engine:} The purpose of the QHE is to produce a high-intensity light field with Poissonian statistics, i.e., laser light.
\item \emph{Light field as battery:} The purpose of the QHE is to increase the light field's ergotropy.
\item \emph{Light field as part of an $N$-partite battery:} The purpose of the QHE is to increase the light field's total ergotropy.
\item \emph{Light field as a free-energy resource beyond the most-energetic available thermal bath:} The purpose of the QHE is to increase the light field's non-equilibrium free energy w.r.t.\ $\Th$.
\end{itemize}

\par

The above scenarios may, in one way or another, differ from the concept of (external) work in classical engine cycles but they all have in common that their respective task pertains to creating a state of the light field that (i)~cannot be generated by directly coupling the cavity field to the two thermal baths and (ii)~constitutes an additional resource, beyond the two thermal baths, that enables to later perform a task that would be impossible to perform solely with the initial resources. We believe that these two properties constitute an operationally-meaningful analogue to the classical concept of ``work'' in fully-autonomous QHEs. Namely, that due to the engine action ``more'' can be done with the resulting piston state than with its initial state, which was assumed to be ``free'' in the sense of thermodynamic resource theories, i.e., a thermal state at one of the two bath temperatures. Phrased differently, the engine allows to generate an out-of-equilibrium state of the cavity field that would be inaccessible solely given the cold and hot thermal baths.

\par

Depending on the chosen task, the efficiency of the engine has to be defined accordingly. This ambiguity may perhaps be unsatisfying but we should keep in mind that the term ``work'' itself carries an inherent operational definition: Work is the ``useful'' energy transfer to the piston which an agent can afterwards use to perform a task and the remainder is ``heat'' in the sense of wasted energy. Consider, for example, an agent that is limited to a certain set of unitaries~\cite{brown2016passivity} which is incompatible with the piston state generated by the engine---the engine would only produce waste energy (``heat'') that cannot be exploited by this agent. A prime example is the Poissonian maser state which cannot be fully exploited with Gaussian operations. An operational definition of work extraction in the sense of classical thermodynamics will always involve some external agent that controls the system~\cite{alicki2017gkls,gallego2016thermodynamic,boes2018statistical,faist2018fundamental,bera2019thermodynamics,boes2019vonneumann,wilming2019equilibration}. Therefore the piston ergotropy is only an upper bound on the actual unitarily-extractable energy. Any restriction of the agent increases the piston's passive energy, i.e., the energetic weight of its entropy.

\par

We first consider an autonomous QHE whose load is the laser field (yellow path in Fig.~\ref{fig_engine}). There is no external agent that further strives to extract work from the latter. Hence, the quantity of interest is the continuously increasing energy (intensity) of the light field, $\dot{E}_\indexf>0$, after the atom has reached a steady state. The ``natural'' efficiency for this task is the energetic efficiency
\begin{equation}\label{eq_eta_energy}
  \eta_\mathrm{ss}^E\coloneq\frac{\dot{E}_\indexf}{\Jh}\leq 1-\frac{\Tc}{\Th}+\frac{\Tc \dot{\mathcal{S}}_\mathrm{af}}{\Jh},
\end{equation}
where $\Jh>0$ is the heat current from the hot bath to the atom and $\dot{\mathcal{S}}_\mathrm{af}>0$ the change in the von-Neumann entropy of the joint atom--field system. Owing to the increase of the latter the energetic efficiency~\eqref{eq_eta_energy} is not bounded by the Carnot efficiency~\cite{boukobza2013breaking}. However, Eq.~\eqref{eq_eta_energy} is not the thermodynamic efficiency of heat-to-work conversion and as such cannot be directly compared to the Carnot bound~\cite{carnotbook} as it pertains to different energetic quantities.

\par

By contrast, if the task of the engine is to autonomously charge a battery from which work is extracted later on in an ideal controlled (non-autonomous) process (green branch in Fig.~\ref{fig_engine}) then not only the intensity of the light matters, but also its quantum statistics and entropy. The corresponding efficiency must then refer to the light field's ergotropy. Under steady-state operation of the engine above threshold the atom already relaxed to a stationary state while the field is in a mixed Poissonian state with a monotonically-increasing intensity and entropy. To fulfill the sub-additivity of entropy, $\mathcal{S}_\mathrm{af}\leq\mathcal{S}_\mathrm{a}+\mathcal{S}_\mathrm{f}$, at any time in this steady-state regime where $\mathcal{S}_\mathrm{a}=\mathrm{const.}$ and $\dot{\mathcal{S}}_\indexf>0$, the total (atom--field) entropy cannot monotonically increase faster than the field entropy, $\dot{\mathcal{S}}_\mathrm{af}\leq\dot{\mathcal{S}}_\indexf$, as eventually it would surpass the sum of the atomic and field partial entropies. Formally, $\dot{\mathcal{S}}_\mathrm{af}$ could be instantaneously greater than $\dot{\mathcal{S}}_\mathrm{f}$. However, a situation where the total atom--field entropy toggles between $\dot{\mathcal{S}}_\mathrm{af}\leq\dot{\mathcal{S}}_\mathrm{f}$ and $\dot{\mathcal{S}}_\mathrm{af}>\dot{\mathcal{S}}_\mathrm{f}$ would be physically incompatible with an atomic steady state. Identifying $\dot{E}_{\mathrm{pas},\indexf}-\Tc \dot{\mathcal{S}}_\indexf$ as the change in the non-equilibrium free energy of the passive field state w.r.t.\ the cold-bath temperature $\Tc$, we find the ergotropic efficiency
\begin{equation}\label{eq_eta_ergotropy}
  \eta_\mathrm{ss}^\mathcal{W}\coloneq\frac{\dot{\mathcal{W}}_\indexf}{\Jh}\leq 1-\frac{\Tc}{\Th}-\frac{\dot{\mathcal{F}}^\indexc_\indexf(\pi)}{\Jh}\leq \eta_\mathrm{Carnot}.
\end{equation}
Since for the steady-state operation of the engine $\dot{\mathcal{F}}^\indexh_\indexf(\pi)\geq 0\Rightarrow\dot{\mathcal{F}}^\indexc_\indexf(\pi)\geq 0$ it follows that the ergotropic efficiency is always limited by the Carnot bound. The piston ergotropy is the closest counterpart to the concept from classical thermodynamics that the work performed by an engine can readily be used, i.e., without any further thermodynamic process.

\par

The ergotropy in Eq.~\eqref{eq_eta_ergotropy} pertains to \emph{local} unitaries applied on the cavity field. If the passive state of the latter is non-thermal, however, \emph{global} operations on more copies of the state (e.g., stemming from multiple engines operated in parallel) can ``unlock'' its bound ergotropy [Eq.~\eqref{eq_W_bound}]. The corresponding efficiency then reads
\begin{equation}\label{eq_eta_ergotropy_tot}
  \eta_\mathrm{ss}^{\mathcal{W}_\mathrm{tot}}\coloneq\frac{\dot{\mathcal{W}}_{\mathrm{tot},\indexf}}{\Jh}\leq 1-\frac{\Tc}{\Th}-\frac{\dot{\mathcal{F}}^\indexc_\indexf(\rho_\mathrm{th}^\pi)}{\Jh}\leq\eta_\mathrm{Carnot},
\end{equation}
where we have identified $\dot{\mathcal{F}}^\indexc_\indexf(\rho_\mathrm{th}^\pi)=\dot{E}_\mathrm{th}-\Tc \dot{\mathcal{S}}_\indexf\geq 0$.

\par

Finally, if the piston mode is understood to be a free-energy resource beyond $\Th$ then the adequate efficiency is 
\begin{equation}\label{eq_eta_free_energy}
  \eta_\mathrm{ss}^\mathcal{F}\coloneq\frac{\dot{\mathcal{F}^\indexh_\indexf}}{\Jh}\leq1-\frac{\Tc}{\Th}-\frac{(\Th-\Tc)\dot{\mathcal{S}}_\indexf(\pi)}{\Jh}\leq\eta_\mathrm{Carnot}.
\end{equation}

\par

Consequently, if an external agent is involved the efficiency does not surpass the Carnot bound. We have summarised the thermodynamic tasks and the associated efficiencies in Table~\ref{table_tasks}.

\par
\begin{table}
  \centering
  \includegraphics[width=\columnwidth]{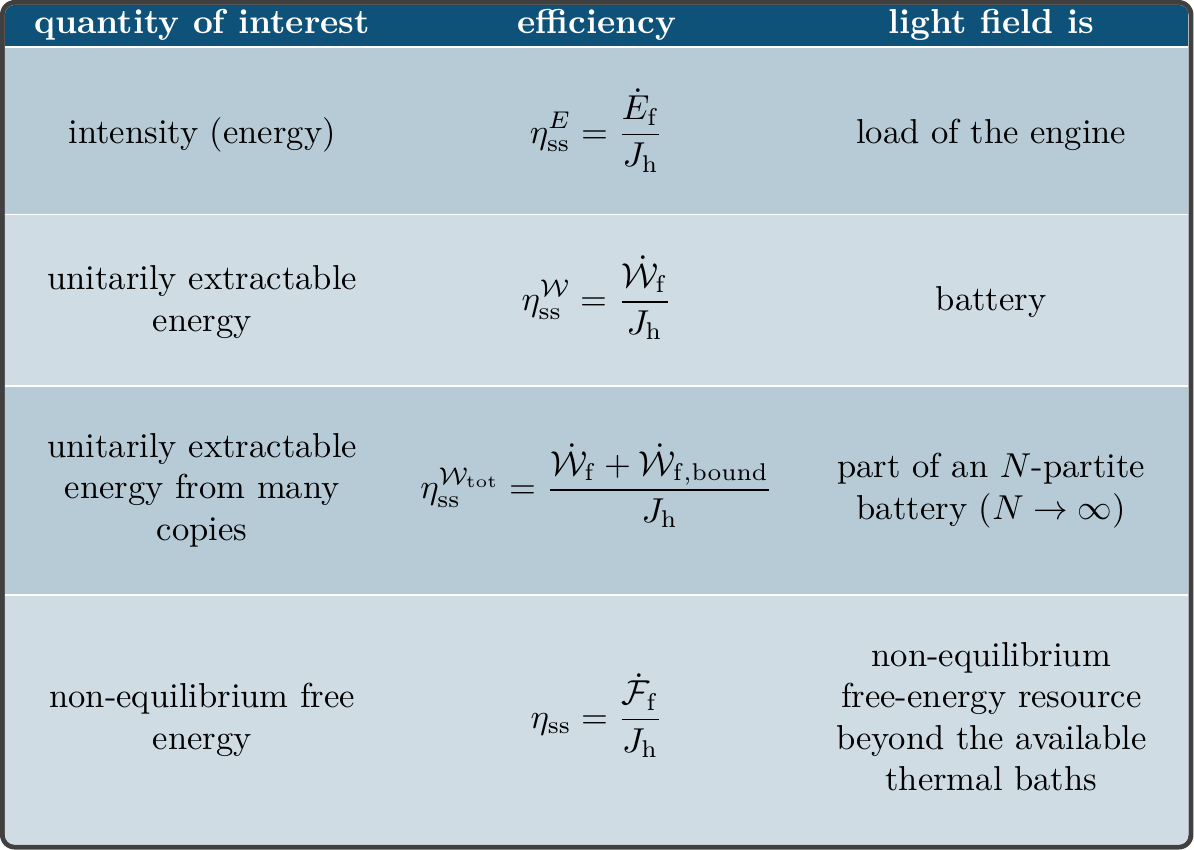}
  \caption{Summary of the thermodynamic tasks and the associated efficiencies of the heat-pumped maser.}\label{table_tasks}
\end{table}
\par
  
Note that the energetic efficiency~\eqref{eq_eta_energy} equals the SSD maser efficiency
\begin{equation}\label{eq_eta_energy_maser}
  \eta^E_\mathrm{ss}=1-\frac{\omegac}{\omegah}\equiv\eta_\mathrm{maser}
\end{equation}
and that in the classical limit of a highly-populated piston mode both the ergotropic and free-energy efficiencies~\eqref{eq_eta_ergotropy} and~\eqref{eq_eta_free_energy} converge towards the latter,
\begin{subequations}\label{eq_eta_convergence}
  \begin{gather}
    \eta^\mathcal{W}_\mathrm{ss}=\eta_\mathrm{maser}\frac{\dot{\mathcal{W}}_{\indexf,\indexss}}{\dot{E}_{\indexf,\indexss}}\approx\eta_\mathrm{maser}\left(1-\sqrt{\frac{2\hbar\omegaf}{\pi E_\indexf}}\right)\\
    \eta^\mathcal{F}_\mathrm{ss}=\eta_\mathrm{maser}\frac{\dot{\mathcal{F}}_{\indexf,\indexss}}{\dot{E}_{\indexf,\indexss}}\approx\eta_\mathrm{maser}\left(1-\frac{\kB\Th}{2 E_\indexf}\right).
  \end{gather}
\end{subequations}
Here we have used the Gaussian approximation of the Poissonian distribution (see Appendix~\ref{app_ehrenfest}). Hence, in the classical limit the energy associated to the field entropy becomes negligible compared to the total field energy. Consequently, one regains the unambiguous notion of work performed by the engine from classical thermodynamics, namely, the entire energy transferred to the piston. This limit, however, does not imply that the operational principle of the maser heat engine becomes classical. Indeed, the maser operation requires discrete energy levels.

\par

The above limit of a classical piston is consistent with M{\o}lmer's argument~\cite{molmer1997quantum} that for large photon numbers the Poissonian distribution becomes so narrow that it may effectively be replaced by a point measure, i.e., a Fock state.

\par

Finally, we note that Eq.~\eqref{eq_eta_energy} is a special case (steady-state operation, $\dot{E}_\mathrm{af}=\dot{E}_\mathrm{f}$) of the general result
\begin{equation}\label{eq_carnot_af}
  \frac{\dot{\mathcal{F}}_\mathrm{af}^\indexc}{\Jh}\leq \eta_\mathrm{Carnot},
\end{equation}
which follows from combining the first and second laws of thermodynamics for the combined atom--field system. Inequality~\eqref{eq_carnot_af} is the equivalent of the famous Carnot formula, which concerns cyclic heat engines, for autonomous engines. Here, work and heat are replaced by the change in the atom--field non-equilibrium free energy w.r.t.\ the cold bath, $\dot{\mathcal{F}}_\mathrm{af}^\indexc$, and the heat current $\Jh$, respectively.

\section{Conclusions}

Heat engines drive the piston mode into an out-of-equilibrium state beyond the available thermal resources (hot and cold baths). Classically, the entropy of the piston is assumed to remain constant, such that all the energy transferred from the engine to the piston is considered to be work. In the quantum domain, however, the entropy change of the piston is no longer negligible and manifests itself by a considerable fraction of the total energy being of passive nature. Hence, different operational notions of work arise, depending on the task of the engine.

\par

We have studied the heat-pumped three-level maser as a simple and illustrative example of an autonomous quantum heat engine (QHE). If the task of this QHE is to drive the light field into a Poissonian state, only the state's energy matters and the energetic efficiency~\eqref{eq_eta_energy} is the adequate performance measure. If, however, the change of the piston state is not the end of the story but an external agent strives to extract work out of the resulting piston state, then the exact task specification and agent capabilities matter. The agent may either unitarily reduce the piston energy, thereby making use of the state's ergotropy, or use its non-equilibrium free energy in a non-unitary process; both leading to different expressions [Eqs.~\eqref{eq_eta_ergotropy} and~\eqref{eq_eta_free_energy}] for the engine efficiency. Here we have reconciled the concepts of (non-equilibrium) free energy---frequently used in statistical mechanics---and ergotropy---a central concept in quantum thermodynamics---in Eqs.~\eqref{eq_F_erg_Fpi} and~\eqref{eq_F_erg_bound_th}.

\par

As revealed by its Poissonian statistics, the entropy of the field generated by the heat-pumped three-level maser solely stems from the random phase~\cite{barnett1986phase,lewenstein1996quantum,molmer1997optical,molmer1997quantum,wiseman1997defining,rudolph2001requirement,vanenk2001quantum,wiseman2002atom,wiseman2003optical,nemoto2004quantum,pegg2005quantum,bartlett2006dialogue,bartlett2007reference,pegg2012physical,wiseman2016how,loveridge2017relativity} and not from heating. Super-Poissonian photon statistics (heating manifested by photon bunching) only occur if the cavity field mode itself is also directly coupled to a thermal bath (cavity decay due to leaky mirrors), which causes this mode to relax to a real steady state with a fixed photon number~\cite{li2017quantum}. The engine then needs to continuously perform work to maintain this out-of-equilibrium state of the light field~\cite{mari2015quantum}.

\par

In the limit of large piston energies (compared to $\hbar\omegaf$ and $\kB\Th$) the ergotropic and free-energy efficiencies [Eqs.~\eqref{eq_eta_ergotropy} and~\eqref{eq_eta_free_energy}] converge towards the Scovil--Schulz-DuBois (SSD) maser efficiency~\eqref{eq_eta_energy_maser}. This may be regarded as the classical limit in which all the energy transferred from the engine to the piston is extractable work, owing to the decreasing relative contribution of passive energy to the total energy. This limit should, however, not be confused with the maser QHE becoming classical---its operation inherently requires discrete energy levels, which is a distinct quantum feature. This classical limit should also be distinguished from the short-time amplifiers considered in Refs.~\cite{gelbwaser2014heat,ghosh2018two} that for large field intensities become classical in the sense that they may be described by controlled, time-dependent external fields even though their working medium remains a quantum object. The steady-state maser considered here, by contrast, is a field generator rather than a field amplifier and hence does not possess a driven counterpart.

\par

While in this work we focused on the quantum state of the piston (field), our results can straightforwardly be extended to the joint atom--field system.

\section*{Acknowledgements} W.\,N.\ acknowledges support from an ESQ fellowship of the Austrian Academy of Sciences (\"OAW). M.\,H.\ acknowledges support by the Austrian Science Fund (FWF) through the START project Y879-N27.

\appendix

\renewcommand\thefigure{\thesection\arabic{figure}}
\setcounter{figure}{0}
\renewcommand{\theequation}{\thesection\arabic{equation}}
\setcounter{equation}{0}

\section{Three-level maser}\label{app_ehrenfest}

The time evolution of the three-level maser in Fig.~\ref{fig_maser}a is governed by the master equation
\begin{equation}\label{eq_master}
  \dot\rho=\frac{1}{i\hbar}\left[H,\rho\right]+\mathcal{L}_\indexh\rho+\mathcal{L}_\indexc\rho
\end{equation}
for the joint atom--field density operator $\rho$~\cite{boukobza2006thermodynamic}. Its coherent part is determined by the Hamiltonian $H=H_\mathrm{free}+\HJC$, which consists of the free part (dropping tensor products with identities on subspaces for notational convenience)
\begin{equation}
  H_\mathrm{free}=\sum_{i=1}^3\hbar\omega_i\proj{i}+\hbar\omegaf a^\dagger a
\end{equation}
and the Jaynes--Cummings interaction~\cite{wallsbook}
\begin{equation}\label{eq_H_JC}
  \HJC=\hbar g\left(\sminus a^\dagger + a \splus\right)
\end{equation}
between the atomic transition $\ket{1}\leftrightarrow\ket{2}$ and the cavity field; here we have defined $\sminus\coloneq\ketbra{1}{2}$ and $\splus\coloneq\ketbra{2}{1}$. The dissipative part of the master equation~\eqref{eq_master} consists of the Liouvillians
\begin{equation}\label{eq_L_h}
  \mathcal{L}_\indexh\rho=\gh(\nh+1)\mathcal{D}\big[\ketbra{1}{3}\big]+\gh\nh\mathcal{D}\big[\ketbra{3}{1}\big]
\end{equation}
and
\begin{equation}\label{eq_L_c}
  \mathcal{L}_\indexc\rho=\gc(\nc+1)\mathcal{D}\big[\ketbra{2}{3}\big]+\gc\nc\mathcal{D}\big[\ketbra{3}{2}\big]
\end{equation}
that describe the coupling of the transitions $\ket{1}\leftrightarrow\ket{3}$ ($\ket{2}\leftrightarrow\ket{3}$) to the hot (cold) thermal bath, respectively, with the dissipator $\mathcal{D}[A]\coloneq 2A\rho A^\dagger-A^\dagger A \rho -\rho A^\dagger A$~\cite{breuerbook}. The thermal populations of the (bosonic) baths are $n_i=\{\exp[\hbar\omega_i/(\kB T_i)]-1\}^{-1}$ ($i\in\{\indexc,\indexh\}$). Note that we do not consider cavity decay since this would involve an additional thermal bath that is directly coupled to the cavity mode; the cavity field would then relax to a steady state with a fixed number of photons.

\par

We have numerically integrated the master equation~\eqref{eq_master} for a large set of different atomic and field parameters. Whereas below the maser threshold the field relaxes to a thermal state, above threshold the field excitation increases while the photon bunching parameter approaches unity, thus revealing the Poissonian statistics. For Fig.~\ref{fig_maser}b we have integrated the master equation~\eqref{eq_master} until $t=100\gh^{-1}$ with the following parameters: $\gc=\gh$, $\omega_1=0$, $\omega_2=\omegaf=30\gh$, $g=5\gh$, $\kB\Tc=20\hbar\gh$ and $\kB\Th=100\hbar\gh$. The maser threshold being $\omega_3=37.5\gh$ we chose $\omega_3=34\gh$ and $\omega_3=150\gh$ to be below and above threshold, respectively. Starting from the atom in the ground state and an empty cavity, at $t=100\gh^{-1}$ the respective photon numbers are $\ew{a^\dagger a}\approx 5.7$ (below threshold) and $\ew{a^\dagger a}\approx 11.7$ (above threshold). All numerical simulations were implemented in Julia using the QuantumOptics.jl framework~\cite{kraemer2018quantumoptics}.

\par

Note that the Liouvillians~\eqref{eq_L_h} and~\eqref{eq_L_c} are of so-called local nature. Namely, they have been derived using local operators, i.e., operators that solely act on one of the two coupled subsystems (atom and cavity), thereby neglecting the atom--field interaction term~\eqref{eq_H_JC}. Here, these local operators describe the atomic transitions $\ket{1}\leftrightarrow\ket{3}$ and $\ket{2}\leftrightarrow\ket{3}$, respectively. While this is a very common strategy in quantum optics~\cite{wallsbook}, the appropriateness and thermodynamic consistency of such local master equations has been debated within the quantum thermodynamics community, see, e.g., Refs.~\cite{levy2014local,hofer2017markovian,gonzalez2017testing}. A conclusion of this debate seems to be that both, the local and the global master equation---the latter being derived from the eigenstates of the interacting system and thus containing global jump operators in the Liouvillian---are adequate in their respective region of validity. We have verified that the second law of thermodynamics (non-negative entropy production) is always fulfilled in our numerical simulations.
  
\par

For the master equation~\eqref{eq_master} the steady-state heat current $\Jh$ from the hot bath to the atom~\cite{alicki1979quantum,kosloff1984quantum} that appears in the efficiencies~\eqref{eq_eta_energy}--\eqref{eq_eta_free_energy} reads
\begin{align}\label{eq_Jh_ss_maser}
  J_{\indexh,\indexss}&\coloneq\Tr\left[H \mathcal{L}_\indexh\rho_\mathrm{ss}\right]\notag\\
  &=\hbar\omegah\big[2\gh\nh P_1^\mathrm{ss}-2\gh(\nh+1) P_3^\mathrm{ss}\big],
\end{align}
where $P_i\coloneq\ew{\proj{i}}$. To evaluate this expression we consider the Ehrenfest equations
\begin{subequations}\label{eq_app_ehrenfest}
  \begin{align}
    \frac{\dd}{\dd t}\ew{a^\dagger a}&=-2 g \imagt\ew{\sigma_+a}\label{eq_app_ehrenfest_aka}\\
    \frac{\dd}{\dd t}P_2&=2 g \imagt\ew{\sigma_+a}+2\gc(\nc+1)P_3-2\gc\nc P_2\label{eq_app_ehrenfest_P2}\\
    \frac{\dd}{\dd t}P_3&=-(2\gh(\nh+1)+2\gc(\nc+1)+2\gh\nh)P_3\notag\\
    &\quad-2(\gh\nh-\gc\nc)P_2+2\gh\nh\label{eq_app_ehrenfest_P3}.
  \end{align}
\end{subequations}
Under steady-state operation of the engine the atomic populations reached their stationary values, $\frac{\dd}{\dd t}P_i^\mathrm{ss}=0$, whilst the photon number in the cavity keeps increasing. Adding Eqs.~\eqref{eq_app_ehrenfest_P2} and~\eqref{eq_app_ehrenfest_P3} when the atom reached steady state and using the normalisation $\sum_{i=1}^3P_i=1$ yields
\begin{equation}
  2 g \imagt\ew{\sigma_+a}_\indexss+2\gh\nh P_1^\mathrm{ss}-2\gh(\nh+1)P_3^\mathrm{ss}=0,
\end{equation}
which, using Eqs.~\eqref{eq_Jh_ss_maser} and~\eqref{eq_app_ehrenfest_aka}, may be recast in the form
\begin{equation}\label{eq_Edot_Jh_ss}
  \dot{E}_{\indexf,\indexss}=J_{\indexh,\indexss}\frac{\omegaf}{\omegah}\equiv J_{\indexh,\indexss}\eta_\mathrm{maser},
\end{equation}
where
\begin{equation}
  \eta_\mathrm{maser}\coloneq \frac{\omegaf}{\omegah}\equiv 1-\frac{\omegac}{\omegah}
\end{equation}
is the SSD maser efficiency~\cite{scovil1959three} and
\begin{equation}\label{eq_Edot_akadot}
  \dot{E}_{\indexf,\indexss}=\hbar\omegaf\frac{\dd}{\dd t}\ew{a^\dagger a}_\indexss.
\end{equation}
Equation~\eqref{eq_Edot_Jh_ss} directly yields the energetic efficiency~\eqref{eq_eta_energy_maser}, which is the SSD maser efficiency.

\par
\begin{figure}
  \centering
  \includegraphics[width=.9\columnwidth]{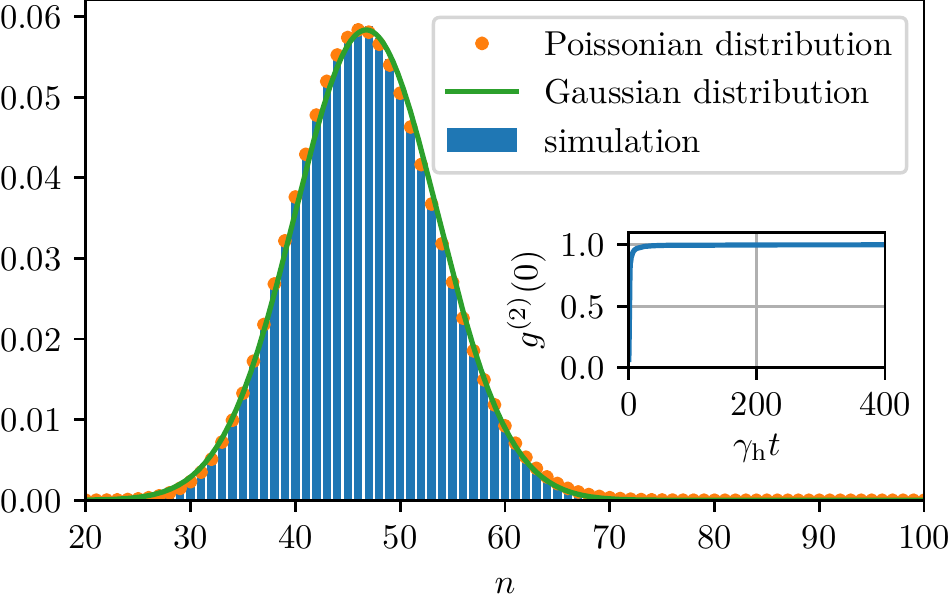}
  \caption{Photon number distribution above threshold (blue bars), Poissonian distribution~\eqref{eq_poisson} with $\alpha^2\coloneq\ew{a^\dagger a}\approx 46.8$ (orange dots) and its Gaussian approximation~\eqref{eq_poisson_gauss} (green curve). Inset: Photon bunching parameter. Same parameters as in Fig.~\ref{fig_maser}b (see text) except that here we have integrated until $t=400\gh^{-1}$ to increase the number of intracavity photons compared to Fig.~\ref{fig_maser}b.}\label{fig_statistics}
\end{figure}
\par

\par

As confirmed numerically (see Fig.~\ref{fig_statistics}), above threshold the reduced intracavity state $\rho_\indexf\coloneq\Tr_\mathrm{a}\rho$ converges towards the Poissonian state 
\begin{subequations}\label{eq_ralpha_together}
  \begin{equation}\label{eq_ralpha}
    \rho_\alpha=\sum_{n=0}^\infty\frac{\alpha^{2n}}{n!}e^{-\alpha^2}\proj{n},
  \end{equation}
  which may equally be decomposed as~\cite{pegg2005quantum,pegg2012physical,wiseman2016how}
  \begin{equation}\label{eq_ralpha_phase_average}
    \rho_\alpha=\frac{1}{2\pi}\int_0^{2\pi}\proj{\alpha e^{i\varphi}}\dd\varphi,
  \end{equation}
\end{subequations}
which may be interpreted as a phase-averaged coherent state. Here we have defined the continuously-growing amplitude
\begin{equation}
  \alpha(t)\coloneq\sqrt{\ew{a^\dagger a}(t)}>0.
\end{equation}
This amplitude should not be confused with a mean field since $\ew{a}=0$. Indeed, the $Q$-function~\cite{wallsbook} of the state~\eqref{eq_ralpha_together} is a rotationally-symmetric annulus devoid of any phase information (Fig.~\ref{fig_maser}b). The state~\eqref{eq_ralpha_together} obeys Poissonian statistics with photon bunching parameter $g^{(2)}(0)=1$. Had we also included cavity decay, i.e., the coupling of the cavity-field mode to the external electromagnetic vacuum, the field would reach a real steady state with super-Poissonian statistics and a fixed photon number~\cite{li2017quantum}. 

\par

We now strive to find analytic expressions for the passive energy and entropy of the laser field state~\eqref{eq_ralpha_together}. To this end we approximate the discrete Poissonian distribution
\begin{equation}\label{eq_poisson}
  p(n)=\frac{\alpha^{2n}}{n!}e^{-\alpha^2}
\end{equation}
that occurs in the state~\eqref{eq_ralpha} by the continuous Gaussian distribution
\begin{equation}\label{eq_poisson_gauss}
  P(n)=\frac{1}{\sqrt{2\pi\alpha^2}}\exp\left(-\frac{\left[n-\alpha^2\right]^2}{2\alpha^2}\right)
\end{equation}
whose expectation value and variance are equal and match their Poissonian counterparts, $\ew{n}_{P}=\var_{P}(n)=\ew{n}_{p}=\var_{p}(n)=\alpha^2$. This approximation follows from the central limit theorem and works well for $\alpha^2\gtrsim 30$ (see Fig.~\ref{fig_statistics})~\cite{tijmsbook}. The energy of this state is $E_\alpha=\hbar\omegaf\alpha^2$ and its passive energy may be computed as
\begin{align}\label{eq_phase_averaged_passive_energy}
  E_{\alpha,\mathrm{pas}}&=\int_0^\infty \hbar\omega_\indexf \left(4n-1\right)\frac{1}{\sqrt{2\pi\alpha^2}}\exp\left(-\frac{n^2}{2\alpha^2}\right)\notag\\
  &=\hbar\omegaf\left(2\sqrt{\frac{2}{\pi}}\alpha-\frac{1}{2}\right).
\end{align}
The entropy of the Poissonian state in the Gaussian approximation~\eqref{eq_poisson_gauss} is
\begin{equation}\label{eq_phase_averaged_entropy}
  \mathcal{S}=\kB\left(\frac{1}{2}+\ln\sqrt{2\pi}+\ln\alpha\right),
\end{equation}
in accordance with Ref.~\cite{scully2017laser}. The non-equilibrium free energy~\eqref{eq_free_energy_noneq} of the Poissonian state~\eqref{eq_ralpha_together} then reads
\begin{equation}\label{eq_F_ralpha}
  \mathcal{F}(\rho_\alpha)=\hbar\omegaf\alpha^2-\kB T\left(\frac{1}{2}+\ln\sqrt{2\pi}+\ln\alpha\right).
\end{equation}

\par

From Eqs.~\eqref{eq_Edot_akadot}, \eqref{eq_phase_averaged_passive_energy} and~\eqref{eq_F_ralpha} then follows
\begin{align}
  \dot{\mathcal{W}}_{\indexf,\indexss}&=\dot{E}_{\indexf,\indexss}-2\sqrt{\frac{2}{\pi}}\hbar\omegaf\frac{\dd}{\dd t}\sqrt{\ew{a^\dagger a}_\indexss}\notag\\
  &=\dot{E}_{\indexf,\indexss}\left(1-\sqrt{\frac{2\hbar\omegaf}{\pi E_{\indexf,\indexss}}}\right)
\end{align}
and
\begin{align}
  \dot{\mathcal{F}}_{\indexf,\indexss}&=\dot{E}_{\indexf,\indexss}-\frac{\kB\Th}{2}\frac{\frac{\dd}{\dd t}\ew{a^\dagger a}_\indexss}{\ew{a^\dagger a}_\mathrm{ss}}\notag\\
  &=\dot{E}_{\indexf,\indexss}\left(1-\frac{\kB\Th}{2E_{\indexf,\indexss}}\right).
\end{align}
Combining these results with Eq.~\eqref{eq_Edot_Jh_ss} then yields the ergotropic and free-energy efficiencies~\eqref{eq_eta_ergotropy} and~\eqref{eq_eta_free_energy}.

\end{document}